\documentclass[prd,twocolumn,aps,amsmath,amssymb,amssymb,nofootinbib,preprintnumbers]{revtex4}

\usepackage{graphicx}
\usepackage{dcolumn}
\usepackage{bm}
\usepackage{amsmath}
\usepackage{amsfonts}

\def\drawbox#1#2{\hrule height#2pt
        \hbox{\vrule width#2pt height#1pt \kern#1pt
              \vrule width#2pt}
              \hrule height#2pt}

\def\Asym#1#2{\vcenter{\vbox{\drawbox{#1}{#2}
              \kern-#2pt       
              \drawbox{#1}{#2}}}}

\newcommand{\be}{\begin{eqnarray}}
\newcommand{\ee}{\end{eqnarray}}
\newcommand{\ba}{\begin{array}}
\newcommand{\ea}{\end{array}}

\begin{document}

\title{Affleck-Dine condensate, late thermalization and the gravitino problem}

\author{Rouzbeh Allahverdi~$^{1}$
and Anupam Mazumdar~$^{2,3}$}
\affiliation{$^{1}$~Department of Physics and Astronomy, University of New
Mexico, Albuquerque, NM 87131, USA.\\
$^{2}$~Physics Department, Lancaster University, Lancaster, LA1 4YB, UK.\\
$^{3}$~Niels Bohr Institute, Blegdamsvej-17, Copenhagen, Denmark.}

\begin{abstract}
In this clarifying note we discuss the late decay of an Affleck-Dine condensate by
providing a no-go theorem that attributes to conserved global charges which are
identified by the net particle number in fields which are included in the flat direction(s). For a rotating condensate, this implies
that: (1) the net baryon/lepton number density stored in the condensate is always
conserved, and (2) the total particle number density in the condensate cannot decrease.
This reiterates that, irrespective of possible non-perturbative particle production
due to $D$-terms in a multiple flat direction case, the prime decay mode of an
Affleck-Dine condensate will be perturbative as originally envisaged. As a result,
cosmological consequences of flat directions such as delayed thermalization as a
novel solution to the gravitino overproduction problem will remain virtually intact.
\end{abstract}

\maketitle


\section{Introduction}

The scalar potential of the Minimal Supersymmetric Standard Model
(MSSM) has many flat directions~\cite{MSSM-REV}. These directions are
classified by {\it gauge-invariant monomials} of the theory, and most
of them carry baryon and/or lepton number~\cite{DRT,GKM}. The flat
directions have many important consequences for the early universe
cosmology~\cite{MSSM-REV}. Most notably, there are two flat directions
which can potentially act as the inflaton and can be tested at the
LHC~\cite{MSSM-INF} (see also~\cite{AKM})~\footnote{Other important
applications of flat directions include curvaton mechasnism~\cite{Curvaton},
inhomogeneous rehetaing and density perturbations~\cite{Inhom}, magnetic field
generation~\cite{AJM}, and non-thermal dark matter~\cite{Q-ball}.}.

Moreover it is well known that a baryon/lepton carrying flat direction
can generate the observed baryon asymmetry via Affleck-Dine
mechanism~\cite{AD}. During inflation a condensate is formed along
the flat direction. After inflation, the condensate starts rotating
once the Hubble rate drops below its mass. This results in a
baryon/lepton asymmetry which will be transferred to fermions upon the
decay of the condensate. The Vacuum Expectation Value (VEV) of the
condensate induces large masses to the fields which are coupled to it.
The decay to these fields will be only possible when the condensate VEV has been
redshifted to sufficiently small values. This will result in a late
perturbative decay of the flat direction condensate~\cite{AD}.
A late decay of an Affleck-Dine condensate has another important
consequence in a supersymmetric universe, namely late thermalization
of the inflaton decay products ~\cite{AVERDI2,AVERDI3}. The
flat direction VEV breaks the Standard Model (SM) gauge
symmetry, thus inducing large masses to gauge (and gaugino) fields
via the Higgs mechanism. This will slow down thermalization by
suppressing dominant reactions which establish kinetic and chemical
equilibrium among the inflaton decay products~\footnote{Finite
temperature effects on MSSM flat directions have been discussed
in~\cite{ROUZBEH,ANISIMOV}.}. A delayed thermalization results in a
reheating temperature much lower than what usually thought. This
naturally solves the outstanding problem of thermal gravitino
overproduction in supersymmetric models~\cite{AVERDI2}~\footnote{Thermalization in the presence of supersymmetric flat directions was first considered in~\cite{EENO} (for some of the works in the non-supersymmetric case, see~\cite{non}). The effect of $Q$-balls formed from fragmentation of flat direction oscillations on reheating is discussed in~\cite{Chung}.}.

The aim of the note is to underline the crucial importance of conserved global charges, which was first observed in
seminal papers by Affleck and Dine~\cite{AD}, and by Dine, Randall and
Thomas~\cite{DRT}. Charges identified by the net particle number
in fields which are included in a flat direction, most notably baryon and lepton number, are preserved by the $D$-terms~\footnote{The $A$-term does
not preserve these charges. However it becomes irrelevant after the very
first oscillations, as it is redshifted away rapidly~\cite{DRT,JOKINEN-THESIS}.}.
For a (maximally) rotating condensate, this implies that possible non-perturbative effects cannot change the baryon/lepton
number density stored in the condensate, and will not decrease the total number
density of quanta in the condensate. As we will briefly mention, under general circumstances, this also holds when $F$-terms are taken into account. The decay of a rotating condensate into other degrees of freedom
happens through the $F$-term couplings. As discussed in the original work
of Affleck and Dine~\cite{AD}, this decay occurs late and is perturbative~\footnote{It is known
that the $F$-terms cannot lead to a non-perturbative decay of a rotating
condensate~\cite{fterm}.}. This guarantees that cosmological consequences of an Affleck-Dine condensate, such as late thermalization, will proceed naturally.

Similar conclusion arises for multiple flat directions represented by
a {\it gauge-invariant polynomial} (for a detailed discussion, see
Ref.~\cite{Enqvist:2003pb,AM}), as it is just a manifestation of the
conservation of global charges carried by a rotating condensate.


\section{Rotating flat directions}

\subsection{Brief introduction to flat directions}

The scalar potential of the MSSM has a large number of flat
directions. The $D$-term and $F$-term contributions to the potential
identically vanish along these directions. The $D$-flat directions are
categorized by gauge-invariant combinations of the MSSM (super)fields
$\Phi_i$. The $D$-flatness requires that~\footnote{Here we use the same
symbol for a superfield and its scalar component.}
\begin{eqnarray} \label{dflatness1}
\sum_{i} {\Phi^*_i T^a \Phi_i} & = & 0 \, \\
\sum_{i} {q_i \vert \Phi_i \vert^2} & = & 0 \, . \label{dflatness2}
\end{eqnarray}
$T^a$ are generators of the $SU(3)_c$ and $SU(2)_L$ symmetries, and
$q_i$ are the charges of $\Phi_i$ under $U(1)_Y$.

A subset of $D$-flat directions are also $F$-flat, in a sense that
the superpotential makes no contribution to the potential at the
renormalizable level. However the flat directions are lifted by
supersymmetry breaking terms, and non-renormalizable superpotential
terms induced by physics beyond the standard model. Hence the
potential along a flat direction, denoted by $\phi$, generically
follows
\begin{equation} \label{flatpotential}
V(\phi) = m^2 \vert \phi \vert^2 + {\lambda}^2 {\vert \phi
\vert^{2(n-1)} \over M^{2(n-3)}} + \Big(A \lambda {\phi^{n} \over M^{n-3}}
+ {\rm h.c.} \Big).
\end{equation}
Here $m,~A \sim {\cal O}({\rm TeV})$ are the soft supersymmetry
breaking mass and $A$-term respectively. $M$ is a high scale where
new physics appears (like $M_{\rm P}$ or $M_{\rm GUT}$), $n > 4$, and
$\lambda \sim {\cal O}(1)$ typically.

The flat direction field $\phi$ acquires a large VEV during inflation
as a result of the accumulation of quantum fluctuations in that
epoch. This leads to the formation of a condensate along the flat
direction~\cite{MSSM-REV}~\footnote{The supergravity corrections to the flat
direction potential are given by $c H^2_{\rm inf} \vert \phi \vert^2$~\cite{DRT}
($H_{\rm inf}$ being the Hubble expansion rate during inflation).
If $0 < c \sim {\cal O}(1)$, the flat direction will settle at the origin during
inflation and will play no dynamical role afterwards. However, if
$0 < c\ll 1$, or for $c < 0$, the flat direction acquires a large VEV during inflation
and plays an important role in the post-inflationary universe. We will be mainly
concerned with the latter case. Note that if there is a positive ${\cal O}(1)$ Hubble
correction to the flat direction mass, then it would also be reflected in  the inflationary potential
which would spoil the success of inflation~\cite{MSSM-REV}.}.
After inflation the VEV of the condensate slowly rolls down. This
continues until the time when the Hubble expansion rate is $\simeq m$,
see Eq.~(\ref{flatpotential}). The VEV of the condensate at this time
is $\varphi_0 \sim (m M^{n-3})^{1/n-2} \gg m$~\cite{DRT}, and all
three terms in Eq.~(\ref{flatpotential}) are comparable in size.

Of particular importance is the $A$-term which, by exerting a torque,
results in the rotation of the condensate. The rotation builds up
already within the first Hubble time~\cite{DRT}. The $A$-term is quickly
redshifted compared with the mass term as it is higher in order (i.e. is
a non-renormalizable term). It becomes negligible after a few
Hubble times, thus leading to a freely rotating condensate. The
trajectory of the motion in the $\phi$ plane is an ellipse, and its
eccentricity is $\simeq 1$ as the $A$-term is initially as large as
the mass term~\cite{DRT}. For our purpose, it can be approximated with
a circular trajectory:
\begin{equation} \label{maximal}
\phi_R = \varphi ~ {\rm cos}(mt) ~ ~ ~ ~ ~ \phi_I = \varphi ~ {\rm sin}(mt),
\end{equation}
where $\varphi$ is redshifted $\propto a^{-1}$ due to Hubble expansion ($a$ being the scale factor of the universe).


\subsection{Physical degrees of freedom}

Let us consider the simplest flat direction represented by the ${H_u
L}$ gauge-invariant combination. Here $H_u$ is the Higgs doublet which
gives mass to the up-type quarks, and $L$ is a left-handed (LH) lepton
doublet~\footnote{The situation will be similar for the $H_u H_d$ flat
direction.}. After imposing the $D$-flatness condition in
Eqs.~(\ref{dflatness1},\ref{dflatness2}), one can always go to a basis where
the complex scalar field (the superscripts denote the
weak isospin components of the doublets and $R,~I$ denote the real and
imaginary parts of a scalar field respectively)
\begin{equation} \label{flat}
\phi = {(H^2_{u} + L^1) \over \sqrt{2}}\,,
\end{equation}
represents a flat direction. The Vacuum Expectation Value (VEV) of $\phi$,
denoted by $\varphi$, breaks the $SU(2)_L \times U(1)_Y$ down to
$U(1)_{\rm em}$~(in exactly the same
fashion as in the electroweak symmetry breaking). The three gauge
bosons of the broken subgroup then obtain masses $\sim g \varphi$ ($g$
denotes a general gauge coupling).

After making the following definitions:
\begin{eqnarray} \label{chi}
\chi_1 & = & {H^2_{u} - L^1 \over \sqrt{2}} \, \\ \nonumber
\chi_2 & = & {H^1_{u} + L^{2*} \over \sqrt{2}} \, \\ \nonumber
\chi_3 & = & {H^1_{u} - L^{2*} \over \sqrt{2}} \, ,
\end{eqnarray}
we find the instantaneous mass eigenstates
\begin{eqnarray} \label{heavy}
\chi^{\prime}_1(t) & = & {{\rm cos}(mt) \chi_{1,R} + {\rm sin}(mt) \chi_{1,I} \over \sqrt{2}} \, \\ \nonumber
\chi^{\prime}_2(t) & = & {{\rm cos}(mt) \chi_2 + {\rm sin}(mt) \chi_3 \over \sqrt{2}} \, ,
\end{eqnarray}
which acquire masses equal to those of the gauge bosons through the $D-$term
part of the scalar potential ($\chi^{\prime}_2$ has both real and imaginary
parts). Note that
\begin{eqnarray} \label{gold}
\chi^{\prime}_3(t) & = & {{\rm cos}(mt) \chi^1_{,I} - {\rm sin}(mt) \chi^1_{,R} \over \sqrt{2}} \, \\ \nonumber
\chi^{\prime}_4(t) & = & {{\rm cos}(mt) \chi_3 - {\rm sin}(mt) \chi_2 \over \sqrt{2}} \, ,
\end{eqnarray}
are the three Goldstone bosons (again $\chi^{\prime}_4$ has both real and imaginary parts),
which are eaten up by the massive gauge fields via the Higgs mechanism.
Therefore, out of the $8$ real degrees of freedom in the two doublets, there are only two {\it
physical} light fields: $\phi_{R}$ and $\phi_{I}$, i.e. the real and
imaginary parts of the flat direction field~\footnote{The masses induced by the flat direction
VEV are supersymmetry conserving. One therefore finds the same mass spectrum in
the fermionic sector as in above (for details, see~\cite{AM}). Note however that scalars also
acquire soft supersymmetry breaking masses $\sim {\cal O}({\rm TeV})$,
while fermions do not. This implies that the fermionic partner of the
flat direction field $\phi$ will remain exactly massless.}.

A rotating flat direction, see Eq.~(\ref{maximal}), does not cross
the origin. Hence, starting with a large VEV such that $g \varphi \gg m$,
the hierarchy between the mass eigenvalues of the heavy and the light
degrees of freedom is preserved at all times. However, rotation results in
time variation in the mass eigenstates of the fields Eq.~(\ref{heavy},\ref{gold}).

The heavy fields, despite having time-varying mass eigenstates~(\ref{heavy}), evolve
adiabatically at all times since $g \varphi \gg m$, and hence will not experience any
non-perturbative effects. In fact, they get decoupled and become dynamically irrelevant.

If there are light fields with a mass $< m$, time variation will become non-adiabatic
and can lead to non-perturbative production of their quanta~\cite{PELOSO,PELOSO2,Basboll:2007vt,Basboll}.
In the case at hand the only physical light degrees of freedom, i.e. degrees of freedom
on the $D$-flat subspace after removing the Goldstone modes~(\ref{gold}), are
the real and imaginary parts of the flat direction field~(\ref{flat}). The mass
eigenstates and eigenvalues of the flat direction are both constant in time, therefore, there
will be no non-perturbative effects whatsoever (for detailed discussion, see Ref.~\cite{AM}).

In general, the number of physical degrees of freedom on the
$D$-flat subspace is given by~\cite{AM}
\begin{equation} \label{lightnum} N_{\rm light} = N_{\rm total} - (2 \times
N_{\rm broken}). \end{equation}
$N_{\rm total}$ is the total number of degrees of freedom in the
gauge-invariant combination which represents the flat direction(s), and
$N_{\rm broken}$ is the number of spontaneously broken symmetries.
The factor $2$ counts for the Goldstone bosons which are eaten by the
gauge fields plus the superheavy degrees of freedom which are decoupled.

In the case of $H_u L$ flat direction, Eq.~(\ref{lightnum}) reads:
$N_{\rm light} = (2 \times 2 \times 2) - (2 \times 3) = 2$, as
explicitly shown above. This is the typical tendency for a single
flat direction in MSSM~\cite{AM}, i.e. when the flat direction is
represented by a gauge-invariant {\it monomial}. Therefore particle
production due to rotation can only be possible if
one considers two or more flat directions, typically represented by a
gauge-invariant {\it polynomial}~\cite{Enqvist:2003pb}.

Nevertheless, we would like to emphasize that time-variation in
the mass eigenstates of light fields, even if it happens, has
practically no bearing for the decay of rotating flat direction(s).
This is the topic which we will discuss in the next section.


\section{No-go theorem for rotating flat directions}

Let us consider MSSM flat direction(s) represented by a gauge-invariant
combination of the fields $\Phi_i$. The $D$-term part of the potential, see
Eqs.~(\ref{dflatness1},\ref{dflatness2}), is invariant under arbitrary phase
transformations of $\Phi_i$ (the same is true for  kinetic terms)~\footnote{Some
combinations of these phases are associated with
the $U(1)_Y$ and $U(1)$ subgroups from diagonal generators of $SU(3)_c$ and
$SU(2)_L$. Hence in a background with non-zero flat direction VEV they
correspond to Goldstone bosons which, in the unitary gauge, are
completely removed from the spectrum. We will deal with this carefully in
the case of explicit examples presented in the next section.}:
\begin{equation}
\Phi_i \rightarrow e^{i \alpha_i} \Phi_i .
\end{equation}
The associated conserved charges (``$\cdot$'' denotes differentiation with respect to time)
\begin{equation} \label{cons}
n_i = i {\dot \Phi^*}_i \Phi_i + {\rm h.c.},
\end{equation}
represent the {\it net} particle number density, i.e. the {\it difference} between the
number density of particles and anti-particles, in $\Phi_i$~\footnote{The $A$-term,
see Eq.~(\ref{flatpotential}), breaks these symmetries. However it virtually decouples
within a few Hubble times after the rotation starts. In particular it will be irrelevant by the
time possible non-perturbative effects become important.}:
\begin{equation}
n_{i}= \left(n_{\rm particle}-n_{\rm anti-particle}\right)_{i}\,.
\end{equation}
The {\it total} particle number density in $\Phi_i$, denoted by ${\tilde n}_i$, is the
{\it sum} of particle and anti-particle number densities:
\begin{equation} \label{inequality}
\tilde n_{i}=\left(n_{\rm particle}+n_{\rm anti-particle}\right)_{i} \geq \vert n_i \vert \,.
\end{equation}
For maximally rotating flat direction(s), the fields $\Phi_i$ follow:
\begin{equation}
\Phi_i = {\phi_i \over \sqrt{2}} ~ {\rm exp}(i \theta_i),
\end{equation}
where ${\dot \phi_i} = 0$, and ${\ddot \theta_i} = 0$ from the equations of motion~\footnote{For a freely rotating scalar field with mass $m_i$ we have ${\dot \theta_i} = m_i$.}. Eq.~(\ref{cons}) then reads
\begin{equation} \label{n}
n_i = {\dot \theta_i} ~ \phi^2_i.
\end{equation}
On the other hand, ${\tilde n}_i$ is given by (the factor of $2$ accounts for the kinetic term and the mass term of $\Phi_i$)
\begin{eqnarray} \label{nf}
{\tilde n}_i = 2 {\vert {\dot \theta}_i \vert}^{-1}
\vert{\dot \Phi}_i \vert^2} = \vert {\dot \theta_i \vert ~ \phi^2_i.
\end{eqnarray}
The inequality in Eq.~(\ref{inequality}) is therefore
saturated, ${\tilde n}_i = \vert n_i \vert$, for maximal rotation.
This is not surprising as such a condensate
consists of particles or anti-particles {\it only}.

This leads to a following {\it no-go theorem}:\\

{\it Consider MSSM flat direction(s) represented by gauge-invariant combinations of fields $\Phi_i$. Possible non-perturbative particle production from time-variation in the mass eigenstates caused by the $D$-terms:\\ \\
\noindent
(1) cannot change the net particle number density in $\Phi_i$, denoted by $n_i$, and hence the total baryon/lepton number density stored in the condensate.\\ \\
\noindent
(2) cannot decrease the total particle number density in $\Phi_i$, denoted by ${\tilde n}_i$, thus the total number density of quanta ${\tilde n}= \sum_{i}{{\tilde n}_i}$ in the condensate~\footnote{Here we mean the comoving quantities as the Hubble expansion inevitably redshifts any physical number density.}. As a direct consequence of the conservation of energy density, non-perturbative effects will not increase the average energy of quanta $E_{\rm ave}$.}\\

We note that, so far as the $D$-terms are concerned, the theorem also applies to the subsequent evolution of the plasma formed after the phase of particle production. This implies that possible non-perturbative effects do not lead to the decay of a rotating condensate. They merely redistribute the energy which is initially stored in the condensate among the fields on the $D$-flat subspace~\footnote{As we will explain later, this is in sharp contrast to non-perturbative particle production from an oscillating condensate, also called preheating, studied in the context of inflaton decay~\cite{PREHEAT}.}.

Some comments are in order before closing this section. Particle production drains energy from the rotating condensate. The question is how this energy transfer affects the flat direction(s) trajectory. The $D$-terms, see Eqs.~(\ref{dflatness1},\ref{dflatness2}), and kinetic terms are invariant under interchanging the real and imaginary components of scalar fields $\Phi_i$ which are included in the flat direction(s). For a circular motion the trajectory itself is invariant under such interchanges. This implies that possible non-perturbative particle production, which is governed by $D$-terms and kinetic terms, will not change the shape of a circular trajectory. Therefore all that happens in the case of maximal rotation, see Eq.~(\ref{maximal}), is a decrease in the radius of the circle $\varphi$.

In reality, the condensate will not undergo maximal rotation and trajectory of its motion will be an ellipse:
\begin{equation} \label{nearmaximal}
\phi_{R} = \varphi ~ {\rm cos} (m t) ~ ~ ~ ~ ~ \phi_I = \alpha \varphi ~ {\rm sin} (m t) ,
\end{equation}
rather than a circle~(\ref{maximal}), where $\alpha < 1$ is a positive number. However, for a generic
initial condition, the rotation will be near maximal $\alpha \sim 0.3$~\cite{DRT}. The difference between the number density of particles and antiparticles in the condensate is given by
\begin{equation} \label{nearcons}
n = 2 \alpha m \varphi^2,
\end{equation}
while the total number density of quanta in the condensate follows
\begin{equation} \label{neartot}
{\tilde n} = (1 + \alpha^2) m \varphi^2 .
\end{equation}
Note that for $\alpha = 1$ the trajectory is a circle and $n = {\tilde n}$ as we discussed above. However, for $\alpha \sim 1$ we have $n \sim {\tilde n}$ (for $\alpha < 1$ we always have ${\tilde n} > n$). The condensate in this case consists mainly of particles (or anti-particles), but it {\it also} contains a small mixture of anti-particles (or particles).
Therefore, in agreement with the conservation of net particle number density, ${\tilde n}$
can in principle decrease by a factor of $r = (1 + \alpha^2)/2\alpha$, such that the small mixture of (anti-)particles will vanish
and consequently ${\tilde n} = n$. For $\alpha \sim 0.3$, we have $r \sim 2$. The possible decrease in ${\tilde n}$ will therefore be of ${\cal O}(1)$. As we will discuss later on, the situation is similar to that for a maximal rotation with regard to the final decay of the flat direction(s) energy density and late thermalization of the universe.

Finally, we note that the $F$-terms, due to quark and lepton mixing, preserve $\sum_{i}{n_i}$
(i.e. the total baryon/lepton number density) instead of each $n_i$. However, $\vert \sum_{i}{n_i} \vert \sim \sum_{i} \vert n_i \vert$, except in some cases where $\sum_{i}{n_i}$ is much smaller than the the individual $n_i$. This requires special initial conditions for which the baryon/lepton number density stored in individual fields are large, but come with opposite signs in such a way that that they conspire to make the total baryon/lepton number density which is stored in the condensate much smaller~\footnote{The only case where this can happen naturally is for the $H_u H_d$ flat direction which carries $B = L =0$. However, this is a single flat direction for which there is no time variation in the light physical degrees of freedom~\cite{AM}. Therefore there can be no non-perturbative particle production in this case at the first place.}. Hence, under general circumstances, the no-go theorem holds when all interactions in the MSSM Lagrangian are taken into account.


\section{Some Examples}

To elucidate the no-go theorem, we consider three representative examples of
MSSM flat directions. Namely, single flat directions consisting of two
and three fields, and multiple flat directions.


\subsection{$H_u L$ direction}

The $D$-terms associated with $SU(2)_L$ and $U(1)_Y$, see
Eqs.~(\ref{dflatness1},\ref{dflatness2}), are invariant under two $U(1)$ symmetries:
\begin{equation}
H_u \rightarrow e^{i \alpha_1} H_u ~ ~ , ~ ~ L \rightarrow e^{i \alpha_2} L,
\end{equation}
and the corresponding charges
\begin{equation} \label{cons1}
n_1 = i {\dot H_u^*} H_u  + {\rm h.c.} ~ ~ ~ ~ n_2 = i {\dot L^*} L + {\rm h.c.},
\end{equation}
are conserved.

In a background of rotating flat direction, transformations generated
by non-diagonal generators of $SU(2)_L$ can be used to situate the VEVs
along $H^2_u$ and $L^1$ (superscripts denote the weak isospin components),
which we denote by $\phi_1$ and $\phi_2$ respectively:
\begin{equation}
\phi_1 = {\varphi \over 2}~{\rm exp}(i \theta_1) ~ ~ , ~ ~ \phi_2 =
{\varphi \over 2}~{\rm exp}(i \theta_2) \,.
\end{equation}
The phase difference $\theta_2 - \theta_1$ is a Goldstone boson which
can be removed through a $U(1)_Y$ transformation~\footnote{Or,
equivalently, the diagonal generator of $SU(2)_L$.} (for identification of
Goldstone modes, see Section IX). This, as shown
before, leaves us with only two light degrees of freedom
\begin{equation} \label{two}
\phi_1 = {\varphi \over 2} ~ {\rm exp}(i \theta) ~ ~ ~ , ~ ~ ~ \phi_2 =
{\varphi \over 2} ~ {\rm exp}(i \theta).
\end{equation}
Eq.~(\ref{cons1}) then results in
\begin{equation} \label{n1}
n_1 = n_2 = {\dot \theta} ~ \varphi^2.
\end{equation}
For a rotating flat direction ${\dot \varphi} =0$. Then from the equations of motion we find
${\dot \theta}^2 = (m^2_H + m^2_L)/2$, where $m_H$ and $m_L$ are the masses of $H$ and $L$ respectively. Note that $n_2$ is the lepton number density stored in the condensate.

The total particle number density in $H_u$ and $L$ (denoted by ${\tilde n}_1$ and ${\tilde n}_2$ respectively) follow from Eq.~(\ref{nf}):
\begin{eqnarray} \label{nf1}
{\tilde n}_1 & = & \vert {\dot \theta} \vert ~ \varphi^2 = \vert n_1 \vert \, , \nonumber \\
{\tilde n}_2 & = & \vert {\dot \theta} \vert ~ \varphi^2 = \vert n_2 \vert \, .
\end{eqnarray}
%


\subsection{$udd$ and $LLe$ directions}

The situation for $udd$ and $LLe$ flat directions is quite similar. We
therefore concentrate on the $udd$ case. The $D$-terms associated with
$SU(3)_c$ and $U(1)_Y$, see Eqs.~(\ref{dflatness1},\ref{dflatness2}), are
invariant under three $U(1)$ symmetries (subscripts are the family indices):
\begin{equation}
u_i \rightarrow e^{i \alpha_1} u_i ~ ~ , ~ ~ d_j \rightarrow e^{i \alpha_2} d_j ~ ~ , ~ ~
d_k \rightarrow e^{i \alpha_3} d_k,
\end{equation}
and the corresponding charges
\begin{equation} \label{cons2}
n_1 = i {\dot u_i}^* u_i  + {\rm h.c.} ~ ~ ~ ~ n_2 = i {\dot d_j}^* d_j + {\rm h.c.} ~ ~ ~ ~
 n_3 = i {\dot d_k}^* d_k + {\rm h.c.} ,
\end{equation}
are conserved.

In a rotating flat direction background, transformations generated by non-diagonal generators of
$SU(3)_c$ can be used to situate the VEVs along $u^1_i$, $d^2_j$, $d^3_k$ (which
we denote by $\phi_1$, $\phi_2$, $\phi_3$ respectively), where superscripts denote the color indices:
\begin{eqnarray}
\phi_1 & = & {\varphi \over \sqrt{6}}~{\rm exp}(i \theta_1)\, , \nonumber \\
\phi_2 & = & {\varphi \over \sqrt{6}}~{\rm exp}(i \theta_2) \, , \nonumber \\
\phi_3 & = & {\varphi \over \sqrt{6}}~{\rm exp}(i \theta_3). \,
\end{eqnarray}
The phase
differences $(2 \theta_1 - \theta_2 - \theta_3)$ and $\theta_1 -
\theta_2$ are Goldstone modes which can be removed through
transformations generated by diagonal generators of
$SU(3)_c$~\footnote{The action of $U(1)_Y$ is the same as that of the ${\bf (-1,-1,+2)}$
diagonal generator of $SU(3)_c$.} (for identification of Goldstone bosons, see Section XI).
After the removal of Goldstone bosons, only two light degrees of freedom
remain~\footnote{This is different from the toy example presented in
Ref.~\cite{Basboll:2007vt}, which considers a flat direction consisting of three
fields charged under a single $U(1)$ gauge symmetry. In the case of MSSM,
there are enough symmetries to rotate away all phase differences among the fields, and
hence only the overall phase remains as a physical degree of freedom.}
\begin{eqnarray} \label{three}
\phi_1 & = & {\varphi \over \sqrt{6}} ~ {\rm exp}(i\theta)
\, , \nonumber \\
\phi_2 & = & {\varphi \over \sqrt{6}} ~ {\rm exp}(i \theta)
\, , \nonumber \\
\phi_3 & = & {\varphi \over \sqrt{6}} ~ {\rm exp}(i \theta) . \,
\end{eqnarray}
For a rotating flat direction we have ${\dot \varphi} = 0$. Then from the
equations of motion we find ${\dot \theta}^2 = (m^2_{u_i} + m^2_{d_j} + m^2_{d_k})/3$.
Eq.~(\ref{cons2}) results in
\begin{equation} \label{n2}
n_1 = n_2 = n_3 = {\dot \theta} ~ \varphi^2.
\end{equation}
Note that $n = n_1 + n_2 + n_3$ is three times the baryon
number density stored in the rotating condensate ($u$ and $d$ have
baryon number $-1/3$).

The total particle number density in $u_i,~d_j,~d_k$ (denoted by
${\tilde n}_1,~{\tilde n}_2,~{\tilde n}_3$ respectively) follow from Eq.~(\ref{nf}):
\begin{eqnarray} \label{nf2}
{\tilde n}_1 & = & \vert {\dot \theta} \vert ~ \varphi^2 = \vert n_1 \vert  , \nonumber \\
{\tilde n}_2 & = & \vert {\dot \theta} \vert ~ \varphi^2 = \vert n_2 \vert  \, , \nonumber \\
{\tilde n}_3 & = & \vert {\dot \theta} \vert ~ \varphi^2 = \vert n_3 \vert \, .
\end{eqnarray}
%


\subsection{$\sum_{i}{H_u L_i}$ multiple flat directions}

Now we consider multiple flat directions represented by the
$\sum_{i=1}^{3}{H_u L_i}$ polynomial where all three $L_i$ doublets have a
non-zero VEV. This case was first considered in Ref.~\cite{AJM}. The
$D$-terms associated with $SU(2)_L$ and $U(1)_Y$, see
Eqs.~(\ref{dflatness1},\ref{dflatness2}), are invariant under four $U(1)$ symmetries:
\begin{eqnarray}
L_1 \rightarrow e^{i \alpha_1} L_1 & , ~ L_2 \rightarrow e^{i \alpha_2} L_2 ~ , & L_3
\rightarrow e^{i \alpha_3} L_3 \, \nonumber \\
& H_u \rightarrow e^{i \alpha_4} H_u & \, ,
\end{eqnarray}
and the corresponding charges
\begin{eqnarray} \label{cons3}
n_1 = i {\dot L_1}^* L_1  + {\rm h.c.} & n_2 = i {\dot L_2}^* L_2 + {\rm h.c.} & n_3 =
 i {\dot L_3}^* L_3 + {\rm h.c.} \, , \nonumber \\
& n_4 = i {\dot H_u}^* H_u + {\rm h.c.} & \, ,
\end{eqnarray}
are conserved.

Similar to the case of the $H_u L$ single flat
direction, the VEVs can be situated along the
$L^1_1,~L^1_2,~L^1_3,~H^2_u$ components which we denote by
$\phi_1,~\phi_2,~\phi_3,~\phi_4$ respectively~\footnote{The situation
is actually more subtle than that for the $H_u L$ single direction, see Section IX.}:
\begin{eqnarray} \label{4}
\, \nonumber \\
\phi_1 & = & {\varphi_1 \over 2} ~ {\rm exp}(i \theta_1)
\, , \nonumber \\
\phi_2 & = & {\varphi_2 \over 2} ~ {\rm exp}(i \theta_2) \, , \nonumber \\
\phi_3 & = & {\varphi_3 \over 2} ~ {\rm exp} (i \theta_3) \, , \nonumber \\
\phi_4 & = & {\varphi \over 2} ~ {\rm exp} (i \theta_4) \, ,
\end{eqnarray}
where $\varphi^2 = \varphi^2_1 + \varphi^2_2 + \varphi^2_3$ is imposed
by the $D$-flatness condition, see Eqs.~(\ref{dflatness1},\ref{dflatness2}).
The phase $(\varphi \theta - \varphi_1 \theta_1 - \varphi_2 \theta_2 - \varphi_3 \theta_3)$
is a Goldstone mode which can be removed by a $U(1)_Y$ transformation~\footnote{Or,
equivalently, a transformation generated by the diagonal generator of
$SU(2)_L$.} (for identification of Goldstone bosons, see Section X). After its
removal we can recast Eq.~(\ref{4}) in the following form~\footnote{Note that if
any two of $\varphi_1,~\varphi_2,~\varphi_3$ are zero, the
situation will be reduced to that for the $H_u L$ single flat direction in
Eq.~(\ref{two}).}:
\begin{eqnarray} \label{four}
\phi_1 & = & {\varphi_1 \over 2} ~ {\rm exp} (i \theta_1) \, , \nonumber \\
\phi_2 & = & {\varphi_2 \over 2} ~ {\rm exp} (i \theta_2) \, , \nonumber \\
\phi_3 & = & {\varphi_3 \over 2} ~ {\rm exp} (i \theta_3) \, , \nonumber \\
\phi_4 & = & {\varphi \over 2} ~ {\rm exp} \Big[i \Big({\varphi_1 \theta_1 +
\varphi_2 \theta_2 + \varphi_3 \theta_3 \over \varphi} \Big) \Big] \, .
\end{eqnarray}
Eq.~(\ref{cons3}) now results in
\begin{eqnarray} \label{number3}
n_1 & = & {\dot \theta_1} ~ \varphi^2_1 \, , \nonumber \\
n_2 & = & {\dot \theta_2} ~ \varphi^2_2 \, , \nonumber \\
n_3 & = & {\dot \theta_3} ~ \varphi^2_3 \, , \nonumber \\
n_4 & = & ({\varphi_1 {\dot \theta_1} + \varphi_2 {\dot \theta_2} +
\varphi_3 {\dot \theta_3}}) ~
 \varphi \, .
\end{eqnarray}
For maximal rotation
$\varphi_1,~\varphi_2,~\varphi_3$ are constant, and ${\ddot \theta_1}
= {\ddot \theta_2} = {\ddot \theta_3} = 0$ from the equations of
motion~\footnote{The situation is more complicated than the previous
examples due to having more than one physical phase. In particular,
$\phi_i$ can rotate with different velocities, where the velocity of
rotation is given by ${\dot \theta_i}$.}.

Note that $n = n_1 + n_2 + n_3$ is the lepton number stored in the condensate.
The total particle number density in $L_1,~L_2,~L_3,~H_u$  (denoted by
${\tilde n}_1,~{\tilde n}_2,~{\tilde n}_3,~{\tilde n}_4$ respectively) follow from Eq.~(\ref{nf}):
\begin{eqnarray} \label{nf3}
{\tilde n}_1 & = & \vert {\dot \theta_1} \vert \varphi^2_1 = \vert n_1 \vert \, , \nonumber \\
{\tilde n}_2 & = & \vert {\dot \theta_2} \vert \varphi^2_2 = \vert n_2 \vert \, , \nonumber \\
{\tilde n}_3 & = & \vert {\dot \theta_3} \vert \varphi^2_3 = \vert n_3 \vert \, , \nonumber \\
{\tilde n}_4 & = & \vert {\varphi_1 {\dot \theta_1} + \varphi_2 {\dot \theta_2} + \varphi_3 {\dot \theta_3}} \vert ~ \varphi = \vert n_4 \vert \, .
\end{eqnarray}
%


\section{Differences between rotating and oscillating condensates}

Let us consider an oscillating condensate for which the trajectory of
motion is a line instead of a circle:
\begin{equation} \label{osc}
\phi_R = \varphi ~ {\rm cos}(mt) \,,~ ~ ~ ~ ~ ~ ~ ~~ ~ ~ \phi_I = 0,
\end{equation}
where $g \varphi \gg m$ ($g$ is a typical gauge coupling).
In this case the mass eigenstate of the $\chi$ fields which are coupled to $\phi$ through the $D$-terms, see Eq.~(\ref{chi}), are constant in time but the mass eigenvalues oscillate. The time variation becomes non-adiabatic as $\phi \approx 0$. As a result, $\chi$ quanta are created within short intervals each time that $\phi$ crosses the origin~\cite{PREHEAT}. This leads to an explosive stage of particle production, also called preheating, which eventually results in a {\it plasma} of $\chi$ and $\phi$ quanta with typical energy
\begin{equation} \label{oscen}
E_{\rm ave} \sim (g \varphi m)^{1/2} \gg m.
\end{equation}
This implies an increase in the average energy of quanta, and hence a decrease
in the number density of quanta, as compared to the original condensate. If the universe
were to fully thermalize after preheating, we would have $E_{\rm ave} \sim T \sim (m \varphi)^{1/2}$. Preheating is therefore a step toward full thermal equilibrium as it partially increases $E_{\rm ave}$ toward irs equilibrium value~\footnote{In reality, it takes a much longer time to establish full thermal equilibrium after preheating~\cite{fk}.}.

This is in sharp contrast to the situation for a rotating condensate. There, as we argued,
possible particle production cannot decrease the number density of quanta. The marked difference between the two cases can be
understood from the trajectory of motion (i.e. circular for rotation versus
linear for oscillation). An oscillating condensate $\phi$ can be
written as
\begin{equation} \label{real}
\phi = {\varphi \over 2} ~ {\rm exp} (i \theta) + {\varphi \over 2} ~ {\rm
exp} (-i \theta),
\end{equation}
and the conserved charge associated with the global $U(1)$
(corresponding to phase $\theta$) is given by
\begin{equation} \label{realn}
n = i {\dot \phi^*} \phi + {\rm h.c.} = 0 .
\end{equation}
This is not surprising since an oscillation is the superposition of two
rotations in opposite directions, which carry exactly the same number of
particles and anti-particles respectively. Therefore the net particle number
density stored in an oscillating condensate is zero.

Now consider non-perturbative particle production from an oscillating condensate.
One can think of this process as a series of annihilations among $N$ particles and $N$
anti-particles in the condensate, $N > 1$, into an energetic particle-anti-particle pair.
This is totally compatible with conservation of charge, see Eq.~(\ref{realn}); $n=0$ after
preheating as well as in the condensate.

On the other hand, a (maximally) rotating condensate
consists of particles or anti-particles {\it only}, see
Eqs.~(\ref{nf1},~\ref{nf2},~\ref{nf3}). Conservation of
the net particle number density then implies that $N \rightarrow 2$
annihilations ($N > 2$) are forbidden: annihilation of particle (or
anti-particle) quanta cannot happen without violating the net particle number
density. Therefore the total number density of quanta will not
decrease, and the average energy will not increase~\footnote{Note that
an increase in the total particle number density, through creation of an
equal number of particles and anti-particles will be in agreement with
the conservation of the net particle number density. In this case the
resulting plasma will be even denser than the condensate.}.


\section{Decay of a rotating condensate}

As we have discussed, any possible non-perturbative particle production will result in a
plasma which is at least as dense as the initial condensate. All that can happen is a redistribution of the energy density in the condensate among the fields on the $D$-flat subspace. These fields have masses comparable to the flat direction mass $m$, as they all arise from supersymmetry breaking. Then, since the average energy is $E_{\rm ave} \leq m$, the resulting plasma essentially consists of non-relativistic quanta. Its energy density $\rho = {\tilde n} E_{\rm ave}$ is therefore redshifted $\propto a^{-3}$ ($a$ is the scale factor of the universe).

The question is when this plasma will decay to {\it other MSSM fields}, in particular fermions, and thermalize. The plasma induces a large mass $m_{\rm eff}$ to the scalars which are not on the $D$-flat subspace and their fermionic partners through the $F$-terms. In the Hartree approximation the effective mass is given by (for example, see~\cite{Tkachev})
%
\begin{equation} \label{meff}
m^2_{\rm eff} \sim h^2 {{\tilde n} \over E_{\rm ave}},
\end{equation}
where $h$ denotes a Yukawa coupling~\footnote{In the case of thermal equilibrium (and zero chemical potential) we have ${\tilde n} \sim T^3$ and $E_{\rm ave} \sim T$, where $T$ is the temperature, yielding the familiar result $m^2_{\rm eff} \sim h^2 T^2$.}. The one-particle decay is kinematically forbidden as long as $m_{\rm eff} \geq  m$. Note that higher order processes such as $N \rightarrow 2$ annihilations ($N > 2$) cannot happen due to conservation of global charges (i.e. baryon and lepton number) in the plasma. Since ${\tilde n}$ and $E_{\rm ave}$ are respectively $\geq$ and $\leq$ than their corresponding values in the initial condensate, thus $m_{\rm eff}$ will always be larger than the induced mass by the condensate VEV, which is given by $h \varphi$.

Further note that $m_{\rm eff}$ is redshifted $\propto a^{-3/2}$, where $a \propto H^{-2/3}~(H^{-1/2})$ in a matter (radiation) dominated epoch. The decay of energy density $\rho$ (initially stored in the condensate) happens {\it only}  when $m_{\rm eff}$ has been redshifted below $m$, at which time the Hubble expansion rate is given by ($\varphi_0$ is the initial VEV of the condensate)~\cite{AVERDI2,AVERDI3}
%
%
%
%
%
\begin{eqnarray} \label{dec}
H_{\rm dec} & \sim & m \left({m \over h \varphi_0}\right) ~ ~ ~ ~ ~ ~ ({\rm matter ~ domination}) \, , \nonumber \\
H_{\rm dec} & \sim & m \left({m \over h \varphi_0}\right)^{4/3} ~ ~ ({\rm radiation ~ domination}) \, .
\end{eqnarray}
Hence for a large $\varphi_0$ the decay time scale is sufficiently large compared to the time scale for possible non-perturbative particle production, which is $\sim m^{-1}$. This decay happens perturbatively as discussed in~\cite{DRT,AD}.

As we discussed earlier, in reality the condensate has an elliptic trajectory whose eccentricity is $\sim 0.3$. This, see Eqs.~(\ref{nearcons},\ref{neartot}), implies that ${\tilde n}$ can at most decrease (and hence $E_{\rm ave}$ increase) by a factor of 2 compared with their corresponding values in the initial condensate. Therefore $m_{\rm eff}$, see Eq.~(\ref{meff}), may be smaller by a factor of 2 in the case of a realistic elliptic trajectory. According to Eq.~(\ref{dec}), this will result in an $H_{\rm dec}$ which is larger by a similar factor factor, thus a slightly earlier perturbative decay. Nevertheless, for $\varphi_0 \gg m$, the final decay happens much later than the initial phase of non-perturbative particle production.

This reiterates the main point of this note: a phase of non-perturbative particle production due to rotation, although possible, cannot lead to the decay of flat direction energy density. It will merely result in a redistribution of energy density on the $D$-flat subspace. The final decay (to other fields) will happen late, and will be perturbative, as originally envisaged~\cite{DRT,AD}.


\section{Cosmological Consequences}

For practical purposes, the resulting plasma will behave the same as the initial condensate. In this section we discuss some of the important cosmological consequences for an Affleck-Dine condensate which gives rise to delayed thermalization and a solution
to the gravitino problem.


\subsection{Delayed thermalization}

The condensate VEV, denoted by $\varphi$, spontaneously breaks the SM gauge symmetry
and induces a large mass for the gauge/gaugino fields $m_{\rm eff}
\sim g \varphi$ via the Higgs mechanism. Such a large mass suppresses gauge interactions which play the main
role in establishing thermal equilibrium among inflaton decay
products~\cite{AVERDI2}. For a rotating condensate $\varphi$
changes only due to the Hubble redshift. The gauge interactions will
therefore remain ineffective for a long time until $\varphi$ has been
redshifted to a sufficiently small value. It is only at this time that
full thermal equilibrium can be established~\cite{AVERDI2,AVERDI3}.

Now consider the plasma consisting of quanta of the fields $\Phi_i$ on the $D$-flat subspace. The SM gauge symmetry is broken in the presence of this plasma as well. This results in an induced mass $m_{\rm eff}$ for the gauge fields (and gauginos) which, as mentioned earlier, is given by
\begin{equation}
m^2_{\rm eff} \sim g^2 {{\tilde n} \over E_{\rm ave}}.
\end{equation}
Since $n$ and $E_{\rm ave}$ are respectively $\geq$ and $\leq$ than the corresponding values in the initial condensate, it turns out that $m_{\rm eff} \geq g \varphi$. This implies that the gauge interactions will be (at least) as suppressed as that in the presence of a condensate. Hence, considering that the plasma decays like the initial condensate, thermalization will also be delayed similarly, for details see~\cite{AVERDI2,AVERDI3}.

Again we note that for a realistic elliptic trajectory $m_{\rm eff}$ may be smaller by a factor of 2. However, for $\varphi_0 \gg m$, universe thermalization will still be considerably delayed relative to an initial phase of non-perturbative particle production from the rotating flat direction(s).


\subsection{Thermal generation of gravitinos}

Late thermalization has an important consequence for thermal production of gravitinos~\cite{Ellis}\footnote{Non-thermal gravitino production at early stages of inflaton oscillations~\cite{Maroto} is not a major issue as discussed in~\cite{Mar}.}.

First, delayed thermalization leads to a considerably low reheat temperature given by the
expression~\cite{AVERDI2}
\begin{equation}
T_{\rm R} \sim \left(\Gamma_{\rm thr} M_{\rm P}\right)^{1/2} \,,
\end{equation}
instead of the usual expression $T_{\rm R} \sim \left(\Gamma_{\rm d} M_{\rm P}\right)^{1/2}$. Here $\Gamma_{\rm thr}$ is the rate for thermalization of the inflaton decay products and $\Gamma_{\rm d}$ is the inflaton decay rate. Suppression of the interactions that lead to establishment of thermal equilibrium, due to the VEV of flat direction(s), implies that $\Gamma_{\rm thr}\ll \Gamma_{d}$~\cite{AVERDI2}, and hence a much lower $T_{\rm R}$ than usually found.

In particular, one can naturally obtain $T_{\rm R} \ll 10^{9}$~GeV which, in the case of a weak scale supersymmetry, is required in order not to distort predictions of the Big Bang Nucleosynthesis by the decay of gravitinos~\cite{Ellis}.

Moreover, before thermalization of the inflaton decay products, scattering processes which lead to gravitino production make a negligible contribution (for details, see~\cite{AVERDI2}). These two effects address the long standing gravitino problem in a natural way within supersymmetry
without invoking any ad-hoc mecahnism.

\section{Conclusion}

The important message of this note is that possible non-perturbative effects stemmed from the $D$-terms have no bearing for the decay of energy density in rotating flat direction(s).
This is due to the conservation of global charges associated with the net particle number density in fields which are included in the flat direction(s), most notably the baryon/lepton number density~\cite{AD,DRT}. For a rotating condensate, this ensures that the total number density of quanta will not decrease and, consequently, the average energy of quanta will not increase.

Thus, in sharp contrast to an oscillating condensate (as in the case of inflaton decay via preheating), all that can happen is a mere redistribution of the condensate energy among the fields on
the $D$-flat subspace. The actual decay into other fields happens perturbatively as originally envisaged by Affleck and Dine~\cite{AD}. This ensures the success of
cosmological consequences such as delayed thermalization as a novel solution to the gravitino problem~\cite{AVERDI2,AVERDI3}.


\section{Acknowledgements}

The authors wish to thank A. Jokinen for valuable discussions and collaboration at earlier stages of this work. The research of A.M. is partly supported by the European Union through
Marie Curie Research and Training Network ``UNIVERSENET''
(MRTN-CT-2006-035863) and STFC (PPARC) Grant PP/D000394/1. A.M. would also
like to thank KITP for kind hospitality and partial support by the NSF-PHY-0551164 grant during the course of this work.



\section{Identification of Goldstone bosons}

Here we quickly comment on identification of the Goldstone bosons and their removal from the spectrum. For simplicity, we consider the case with a single $U(1)$ gauge symmetry, but generalization to non-abelian symmetries is straightforward.

Consider $n$ scalar fields $\phi_i$ with respective charges $q_i$ ($1 \leq i \leq n$) under the $U(1)$ symmetry. Covariant derivatives of the scalar fields are
\begin{equation} \label{kinetic}
\sum_{i=1}^{n} {(\partial^\mu + i A^\mu) \phi^*_i ~ (\partial_\mu - i A_\mu) \phi_i}.
\end{equation}
The scalar fields can be written in terms of radial and angular components (denoted by $\varphi$ and $\theta$ respectively):
\begin{equation} \label{polar}
\phi_i = {\varphi_i \over \sqrt{2}} ~ {\rm exp} (i \theta_i).
\end{equation}
Expanding the fields around a background where $\varphi_i$ is constant, as happens for rotating flat direction(s), we then have
\begin{equation} \label{gold1}
\partial_\mu \phi_i = i ({\partial_\mu \theta_i}) ~ {\varphi_i \over \sqrt{2}} ~ {\rm exp}(i \theta_i).
\end{equation}
It can be seen that the combination $\sum_{i=1}^{n}{(\varphi_i q_i \theta_i)}$ can be eliminated from Eq.~(\ref{kinetic}) by performing the following gauge transformation:
\begin{eqnarray} \label{gauge}
\theta_i & \rightarrow & \theta_i + q_i \theta_i \, , \nonumber \\
A_\mu & \rightarrow & A_\mu - \sum_{i=1}^n {q_i (\partial_\mu \theta_i)} \, .
\end{eqnarray}
Therefore it is not a true physical degree of freedom. This particular combination is nothing but the Goldstone boson from spontaneous breaking of $U(1)$ symmetry by non-zero values of $\varphi_i$.


\section{$\sum_i {H_u L_i}$ multiple flat directions}

Consider a general VEV configuration of $H_u$ and $L_i$ ($1 \leq i \leq 3$) that satisfies the $D$-flatness condition in Eqs.~(\ref{dflatness1},\ref{dflatness2}). We can always use non-diagonal generators of $SU(2)_L$ to rotate $H_u$ to a basis where $\langle H^1_u \rangle = 0$ (superscripts denote the weak isospin component).

In the case of $H_u L$ single flat direction (where two of the $L_i$ have zero VEV), $D$-flatness under the non-diagonal generators directly implies $\langle L^2 \rangle = 0$ in this basis. However, for multiple flat directions, it is not so obvious that $\langle L^2_1 \rangle = \langle L^2_2 \rangle = \langle L^2_3 \rangle = 0$ in the basis where $\langle H^1_u \rangle = 0$.

Hence let us consider a general configuration where both isospin components of $L_i$ have a non-zero VEV. The vanishing of the $D$-term associated with the diagonal generator of $SU(2)_L$ then implies
\begin{equation} \label{d1}
- \vert \langle H^2_u \rangle \vert^2 + \sum_{i=1}^{3} {\vert \langle L^1_i \rangle \vert^2} - \sum_{i=1}^{3} {\vert \langle L^2_i \rangle \vert^2} = 0,
\end{equation}
while vanishing of the $D$-term associated with $U(1)_Y$ requires that (note that $H_u$ and $L_i$ have opposite hypercharge quantum numbers)
\begin{equation} \label{d2}
\vert \langle H^2_u \rangle \vert^2 - \sum_{i=1}^{3} {\vert \langle L^1_i \rangle \vert^2} - \sum_{i=1}^{3} {\vert \langle L^2_i \rangle \vert^2} = 0.
\end{equation}
It is readily seen that the third term on the LH side of Eqs.~(\ref{d1},\ref{d2}) must vanish, thus $\langle L^2_1 \rangle = \langle L^2_2 \rangle = \langle L^2_3 \rangle = 0$. Also:
\begin{equation} \label{d3}
\vert \langle H^2_u \rangle \vert^2 = \sum_{i=1}^{3} {\vert \langle L^1_i \rangle \vert^2}.
\end{equation}
%


\end{document}